\documentclass[a4paper,12pt]{article}

\pdfoutput=1

\usepackage[paper=letterpaper,margin=1in]{geometry}

\usepackage{amsmath,amssymb,amsfonts,epsfig,cite,setspace,url,color}

\usepackage{hyperref,caption,setspace}

\usepackage{pdfpages,lipsum}

\begin{document}

%%%%%%%%%%%%%%%%%%%%%%%%%%%%%%%%%%%%%%%%%%%%%%%%%%%%%%%%%%%%%%%%%%

%\includepdf[pages={12-15,23,45-49}]{main.pdf}

\vskip 0.25in

\newcommand{\nn}{\nonumber}
\def\cD{{\mathcal D}_C }
\newcommand{\comment}[1]{}

\comment{
% make bibliography single-spaced
\let\oldthebibliography=\thebibliography
\let\endoldthebibliography=\endthebibliography
\renewenvironment{thebibliography}[1]{%
\begin{oldthebibliography}{#1}%
\setlength{\parskip}{0ex}%
\setlength{\itemsep}{0ex}%
}%
{%
\end{oldthebibliography}%
}
}

\def\theequation{\thesection.\arabic{equation}}
\newcommand{\setall}{\setcounter{equation}{0}
        \setcounter{theorem}{0}}
\newcommand{\setequation}{\setcounter{equation}{0}}
\renewcommand{\thefootnote}{\fnsymbol{footnote}}

~\\
\vskip 1cm

\begin{center}
{\Large \bf  A Visualization of the Classical Musical Tradition}

\medskip
\vspace{4mm}

{\large Yang-Hui He$^{1,2,3}$}

\vspace{1mm}

\renewcommand{\arraystretch}{0.5} 
{\small
{\it
\begin{tabular}{rl}
  ${}^{1}$ &
  Merton College, University of Oxford, OX14JD, UK\\
  ${}^{2}$ &
  Department of Mathematics, City, University of London, EC1V 0HB, UK\\
  ${}^{3}$ &
  School of Physics, NanKai University, Tianjin, 300071, P.R.~China\\
\end{tabular}
}
~\\
~\\
~\\
hey@maths.ox.ac.uk
}
\renewcommand{\arraystretch}{1.5} 

\end{center}

\vspace{10mm}

\begin{abstract}
A study of around 13,000 musical compositions from the Western classical tradition is carried out, spanning 33 major composers from the Baroque to the Romantic, with a focus on the usage of major/minor key signatures. A 2-dimensional chromatic diagram is proposed to succinctly visualize the data. The diagram is found to be useful not only in distinguishing style and period, but also in tracking the career development of a particular composer.
\end{abstract}

%\end{titlepage}

\newpage

\section{Introduction}

Within the Western tradition of music, spanning at least the high Renaissance/early Baroque till the advent of the modern 12-tone paradigm, the choice of key signature is one of the decisive elements to any musical composition, reflecting the composer's mood, influencing the style and determining the basic tonality \cite{white,cohn}. That the statistics of keys should be studied for a composer, much in the spirit of stylo-statistics of analysing a writer's preference of words \cite{herdan}, actually goes back to the beginning of the 20th century \cite{corder}, before text statistics rose to prominence in the 1960s. Today, in the age of information, data analyses on readily electronically accessible oeuvres of artists have become the norm in musical and literary theory.

Whilst there have been statistical investigations on the key signatures for specific composers (e.g. \cite{nett}), comparisons thereof (e.g. \cite{mm}), or on establishing online resources (e.g. \cite{web,kadish}), a unified, analytical framework is still lacking. This is unsurprising given that one cannot seemingly do much more than pair-wise correlative studies between the percentages of key usages between composers
(some nice preliminary work in percentages and visualization of key choices of certain and some aggregate classical composers have been done in \cite{mm,kadish}).
The purpose of this paper is to establish a 2-dimensional plot, which will be called the “chromatic diagram” whose geometry (i) underlies a common ground for comparative musicology; (ii) tracks the evolution of a composer’s preference over his/her career; and (iii) provides a visual classification of style.

It is common knowledge that Western tonal composition in the classical tradition falls into the dichotomy of major/minor, each of which is dictated by a key governed by number of accidentals (sharps and flats) resulting in a circle of fifths (q.~v.~Table \ref{t:k}). 
A natural grading which shall be called degree (to borrow terms from algebraic geometry) is clearly to have plus/minus for the number of sharps/flats, so that C-Major/a-minor is 0, G-Major/e-minor is 1, F-Major/d-minor is $-1$. 

The usual notation of using capitals for major keys and lower-cases for the minor keys is also adopted, whereupon a single letter with $^\sharp$ and $^\flat$ will unambiguously specify the key. Tonally, the cycle of fifths should be comprised of no more than 12 keys. However, though C$^\sharp$ is the same as D$^\flat$ tonally, they do reflect a different compositional mood, so these will be kept as distinct. Nevertheless, keys beyond C$^\sharp$ and C$^\flat$ are extremely rare. In other words, for practical purposes, the need beyond degree $\pm 7$ is unnecessary. In summary, for a given key $k$ of a composition, the degree is defined as 
\begin{equation}
d(k) := |\sharp(k)|- |\flat(k)| \ , 
\end{equation}
where $\left|~\right|$ counts the number of accidentals.

%%%%%%%%%%%%%%%%%%
\begin{table}
\begin{tabular}{|c||c|c|c|c|c|c|c|c|c|c|c|c|c|c|c|c|c|} \hline
Major & 
\ldots & C$^\flat$ & G$^\flat$ & D$^\flat$ & A$^\flat$ & E$^\flat$ & B$^\flat$  & F & C & G & D & A & E & B & F$^\sharp$ & C$^\sharp$ & \ldots
\\ \hline
Minor & 
\ldots &  a$^\flat$ & e$^\flat$ & b$^\flat$ & f & c & g  & d & a & e & b & f$^\sharp$ & c$^\sharp$ & g$^\sharp$ & d$^\sharp$ & a$^\sharp$&\ldots
\\ \hline \hline
Degree & \dots & -7 & -6 & -5 & -4 & -3  & -2 & -1 & 0 & 1 & 2 & 3 & 4 & 5 & 6 & 7 & \ldots \\
\hline        
\end{tabular}
\caption{{\sf
The degree for the number of flats/sharps, organized according to relative major/minors, each going in cycles of 5th.
Tonally, for example F$^\sharp$ is the same as G$^\flat$ though traditionally composers make a distinction between these two.
However, keys beyond C$^\sharp$ and C$^\flat$ are extremely rare.}
\label{t:k}
}
\end{table}
%%%%%%%%%%%%%%%%%%%%

\section{The Chromatic Diagram}
Since the duality between major and minor is so manifest in determining the composition, it is expedient to place the two on perpendicular axes which can be called major and minor (ironically terms also already taken by algebraic geometry in the study of conic sections). 
A two-dimensional lattice - which is doubly-periodically identified - is thus naturally created. This construction is similar in spirit to Euler's Tonnetz \cite{cohn,euler}, though not identical to it. 

%%%%%%%%%%%%%%%%%%%%
\begin{figure}[h!]
$
\begin{array}{cc}
(a)
\begin{array}{c}\includegraphics[trim=0mm 0mm 0mm 0mm, clip, width=3in]{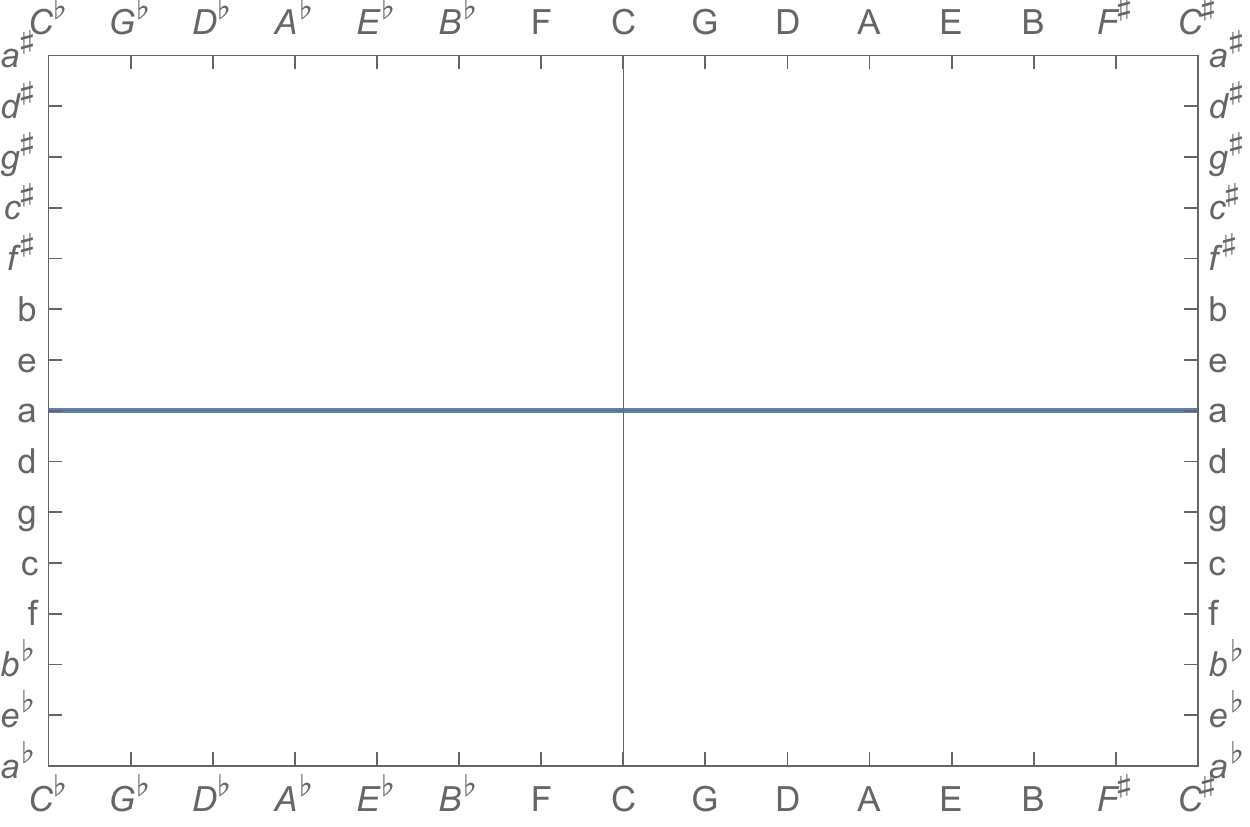}\end{array}
(b)
\begin{array}{c}\includegraphics[trim=0mm 0mm 0mm 0mm, clip, width=3in]{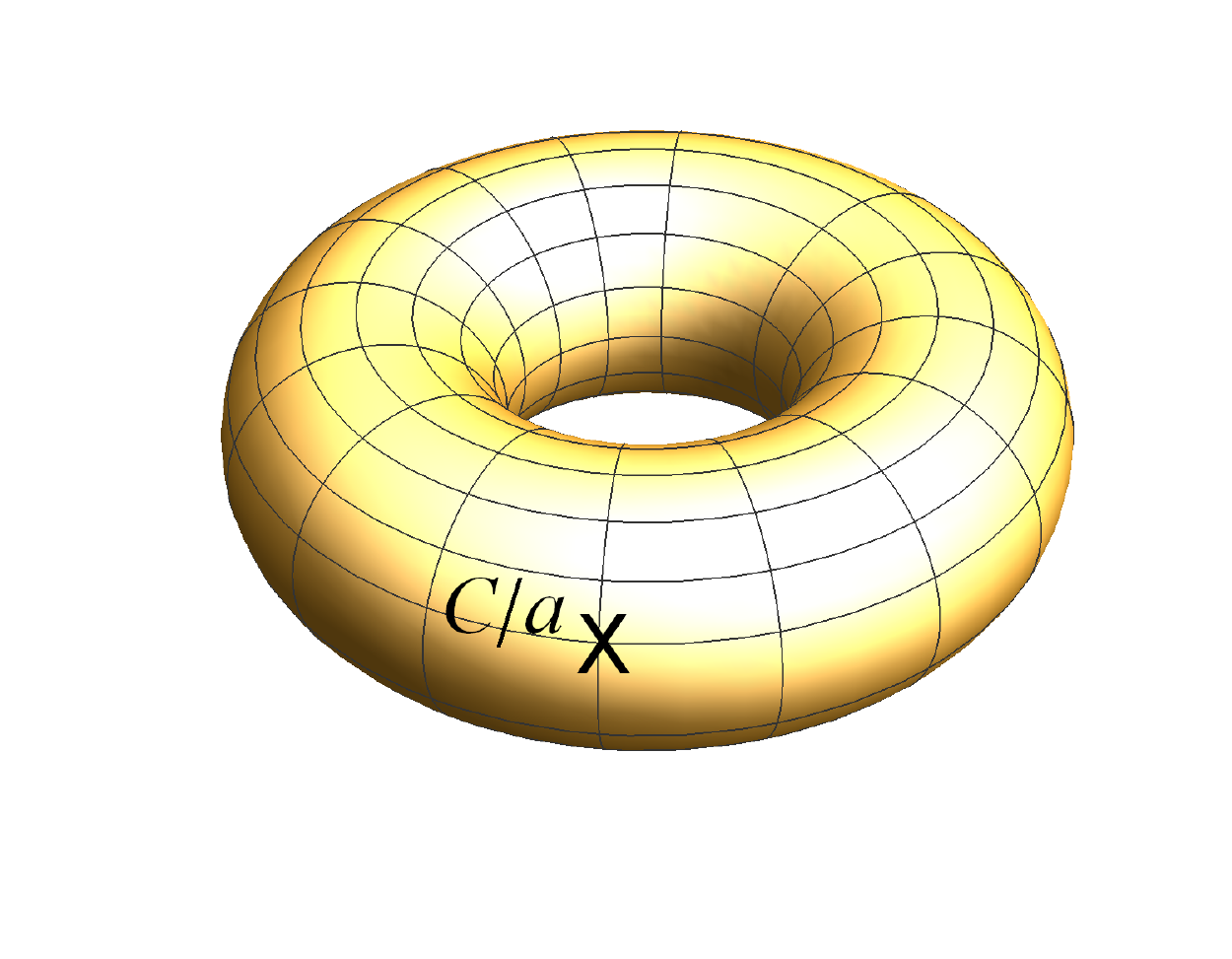}\end{array}
\end{array}
$
\caption{{\sf
(a) The chromatic diagram $\cD$ as a 2-dimensional plot with $(0,0)$ as C-Major/a-minor;
(b) $\cD$ with the cycles of fifths taken into account, giving the topology of a torus with a marked point.
\label{f:D}
}}
\end{figure}
%%%%%%%%%%%%%%%%%%%%%

Now, whilst this definition of the diagram seems to only consider lattice points given by the degree,  averages $\overline{d(k)}$ (both weighted and unweighted) will be shortly taken over a composer’s career, thus non-lattice points will also be occupied. 
Geometrically, as usage of the entire plane and not just the lattice points will be made, this diagram has the topology of a torus $T^2 \simeq S^1 \times S^1$ with a marked point at $(0,0)$ corresponding to C/a.  
This chromatic diagram, henceforth denoted as $\cD$, is illustrated in Figure \ref{f:D} (all plots and computations are done in Mathematica \cite{math}),
both in doubly-periodic planar fashion in part (a) and as the torus in part (b).
 
In summary, points on the chromatic diagram are certain averaged degrees, so that each point corresponds to either a particular composer, to a chosen period of a composer active career, both of which will be explored shortly, 
\begin{equation}
\cD := \{(x,y) = \left( \overline{d(k)}_{k \in \mbox{Major}} , \quad \overline{d(k)}_{k \in \mbox{Minor}} \right) \} \simeq T^2 \ .
\end{equation}

The works of a composer can thus be divided into a list of major (respectively minor) pieces and then averaged (with weights to be discussed shortly) in order to produce a point on $\cD$. 
The complete works of most classical composers are freely available, for instance detailed in \cite{imslp}.
All composers whose catalogue consists of a substantial number of compositions (say around 50 or above) with definitive keys are extracted, which amounts to the list of 33 composers (Table \ref{t:comp}), with a total of 12,331 works, 8,488 of which are in major keys and 3,843 in minors. 

%%%%%%%%%%%%%%%%%%%%
\begin{table}[h!!!]
{\small
$
\begin{array}{|ccccccc|}\hline
 \text{Albinoni}^1 & \text{Bach}^2 & \text{BachCPE}^3 & \text{BachJC}^4 & \text{Beethoven}^5 &
   \text{Brahms}^6 & \text{Bruckner}^7 \\
 \text{Buxtehude}^8 & \text{Chopin}^9 & \text{Clementi}^{10} & \text{Donizetti}^{11} &
   \text{Dvorak}^{12} & \text{Faure}^{13} & \text{Glazunov}^{14} \\
 \text{Handel}^{15} & \text{Haydn}^{16} & \text{Hummel}^{17} & \text{Liszt}^{18} &
   \text{Mendelssohn}^{19} & \text{Mozart}^{20} & \text{Pachelbel}^{21} \\
 \text{Paganini}^{22} & \text{Pleyel}^{23} & \text{Rachmaninoff}^{24} &
   \text{Saintsaens}^{25} & \text{Scarlatti}^{26} & \text{Schubert}^{27} &
   \text{Scriabin}^{28} \\
 \text{Shostakovich}^{29} & \text{Smetana}^{30} & \text{Tchaikovsky}^{31} &
   \text{Telemann}^{32} & \text{Vivaldi}^{33} &  &  \\
\hline
\end{array}
$
}
\caption{{\sf
A total of 33 composers with compositions in distinctly catalogued keys (many with opus numbers); the superscript ordering is maintained throughout. 
}
\label{t:comp}
}
\end{table}
%%%%%%%%%%%%%%%%%%%%

It should be remarked that 
(i) this is a sizable data-set; (ii) some famous composers such as Wagner or Puccini are not included since operas are not usually associated with a particular key; (iii) instrument-specific composers such as Wieniawski (violin) or Sor (guitar), though well-known, are not included because their compositions are biased by the tuning of the instrument (Paganini is included for reference and also because he composed widely for 2 differently tuned instruments); (iv) Renaissance and modern composers whose keys are often modulated at will or are atonal are not included.

\subsection{Comparative Studies and Classification}
\begin{figure}[h!!!]
$
\begin{array}{cc}
(a)
\includegraphics[trim=0mm 0mm 0mm 0mm, clip, width=3in]{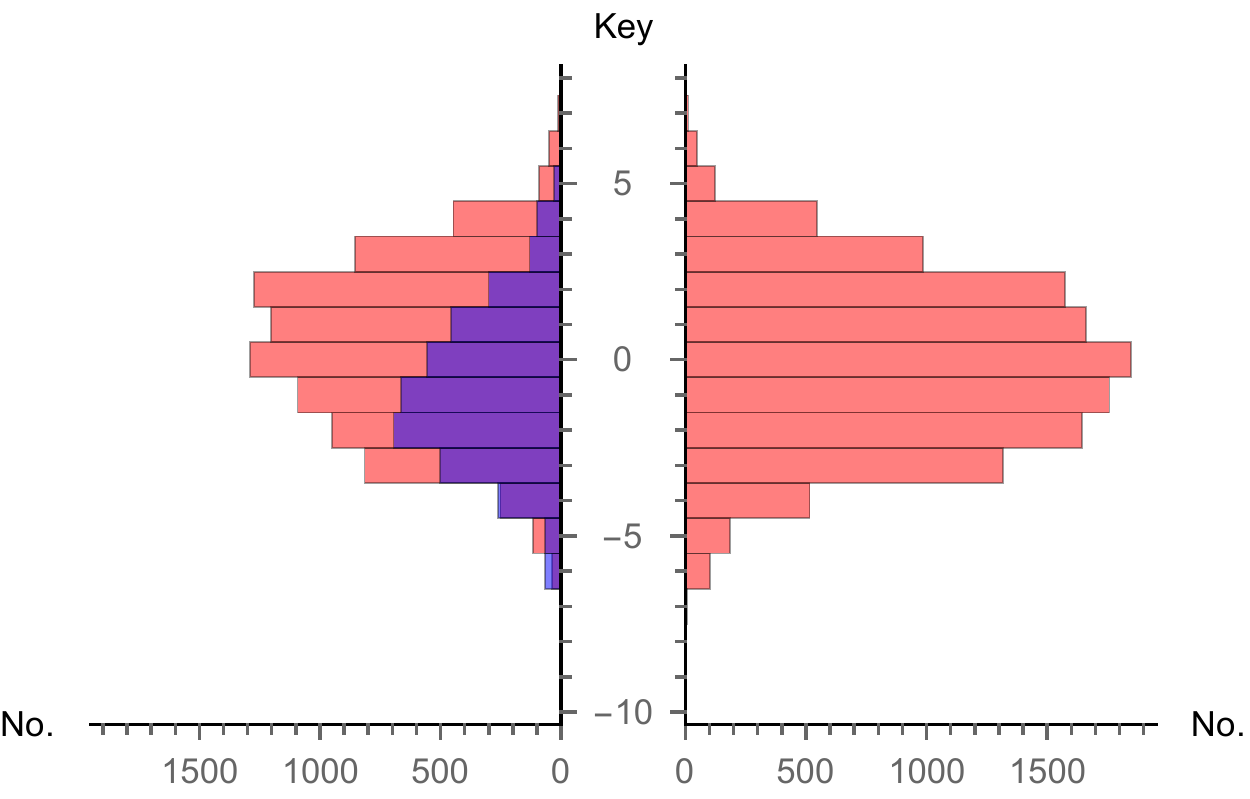}
(b)
\includegraphics[trim=0mm 0mm 0mm 0mm, clip, width=3in]{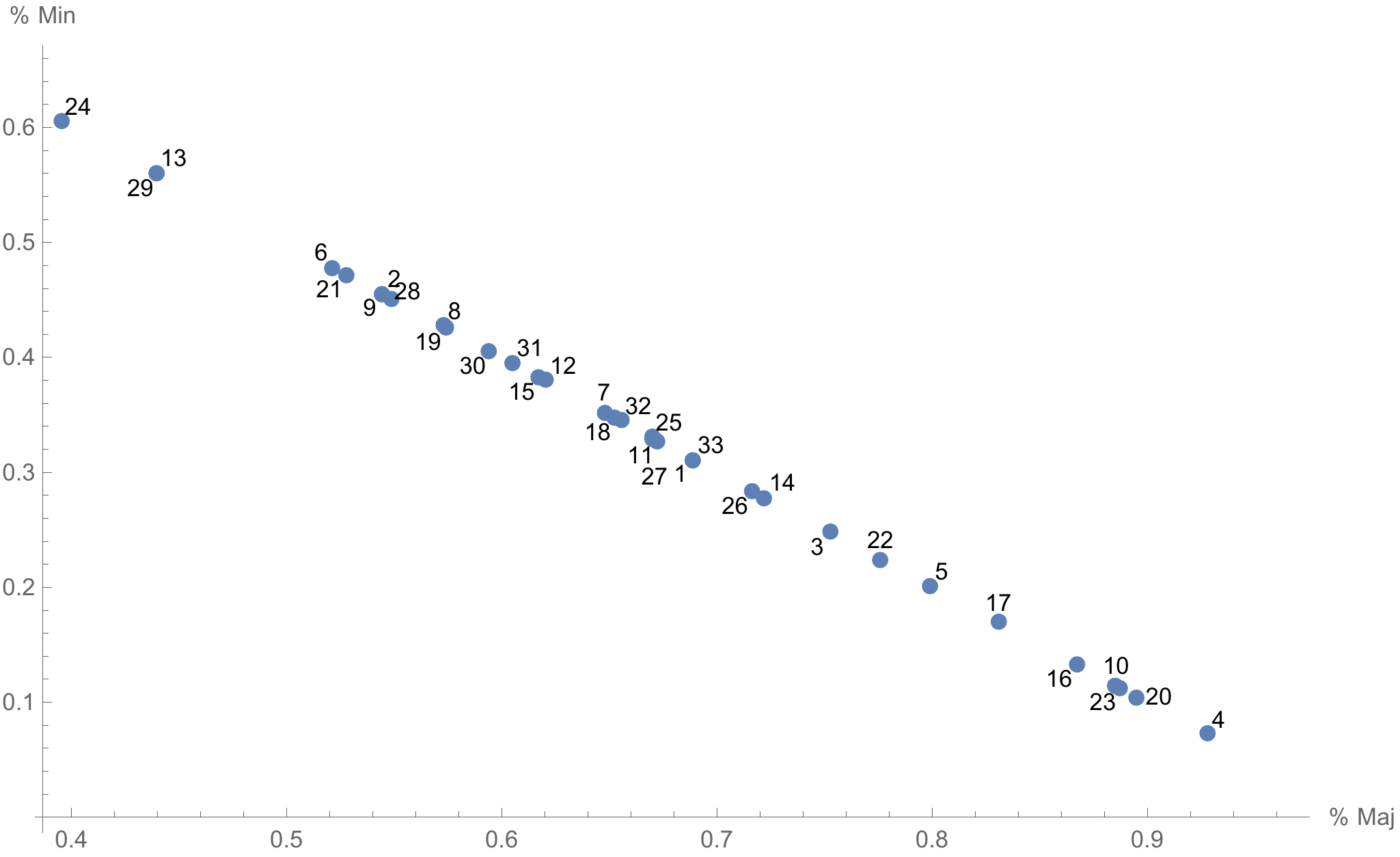}
\end{array}
$
\caption{{\sf
(a) A histogram of major/minor (red/blue) keys over all the works considered, aggregated over all composers on the left and the combined of the two on the right;
(b) The fraction of Major and Minor compositions for the 33 composers (numbered in the box above);
	all points lie on the $x+y=1$ line by construction. 
}
\label{f:hist}}
\end{figure}

First, a histogram of the major and minor compositions, arranged by degree, and collected over all the composers is presented on the left of Figure \ref{f:hist}(a); for reference, the combined histogram is also presented to the right. It is visible that the major key peaks at 0, and the minor, at $-2$, meaning that C and g are the most popular. 
While they appear as normally distributed, tests such as Cramer-Von Mises \cite{math} reject this hypothesis, as they also rejects some of the standard distributions such as Cauchy or Poisson. 
It would be interesting to find an analytic PDF for this data-set, though this is not needed for present analyses.

%%%%%%%%%%%%%%%%%%%%

A scatter plot of the fraction of major (horizontal axis)/minor(vertical axis) compositions is shown in Figure \ref{f:hist}(b); 
all points lie on the $x+y=1$ line by construction but the closeness to either axis indicates a preference of mode. For example, Mozart and 
J.~C.~(the ``English'') Bach clearly prefer major keys while Rachmaninoff and Shostakovich prefer minor; the degree of preference can be measured by, say, the ratio between the $x$ and $y$ intercepts.

\begin{figure}
$
\begin{array}{cc}
(a)
\includegraphics[trim=0mm 0mm 0mm 0mm, clip, width=3in]{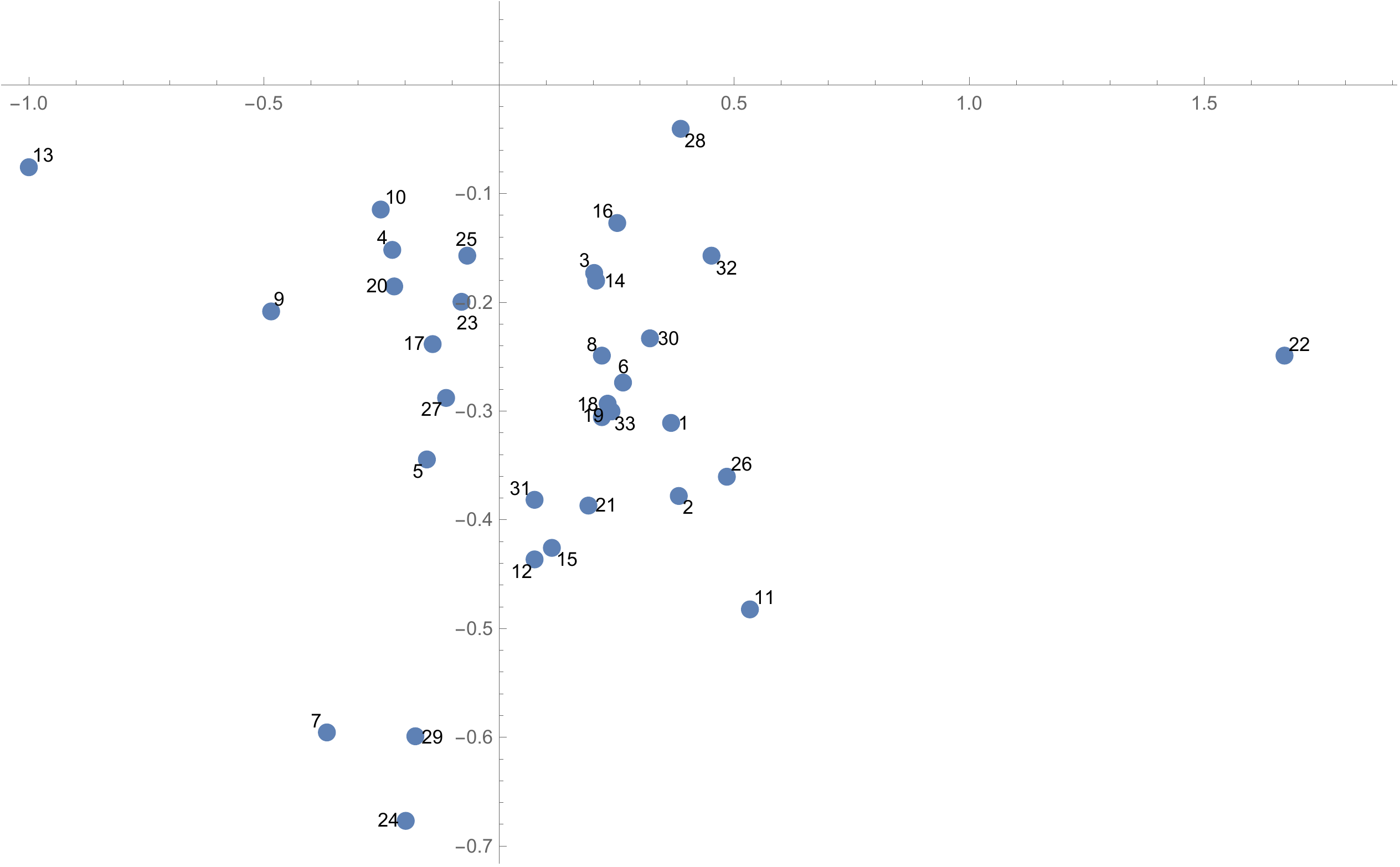}
(b)
\includegraphics[trim=0mm 0mm 0mm 0mm, clip, width=3in]{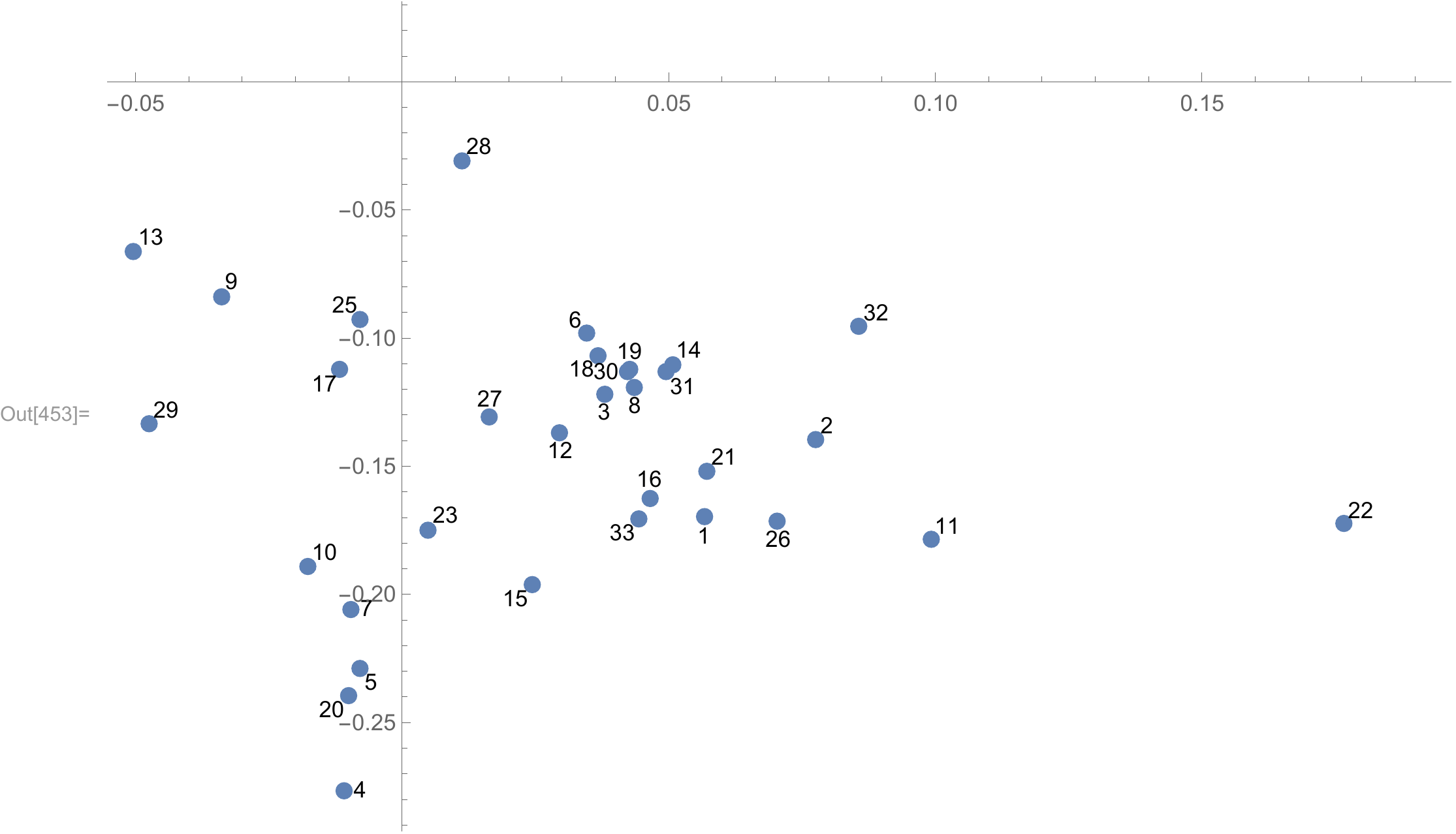}
\end{array}
$
\caption{{\sf
(a) Scatter plot of arithmetic mean of the degrees of all 33 composers' major and minor compositions;
(b) likewise for the weighted (by overall aggregate frequency) means.
}
\label{f:avg}}
\end{figure}

An aggregate view can be seen (Figure \ref{f:avg}(a)) by plotting the arithmetic mean over each artist's major/minor compositions and shown labelled on the chromatic diagram; i.e., 
\begin{equation}
\cD \ni (x_i, y_i) = 
T_i^{-1}
\left(
\sum_{k \in \mbox{{\tiny Maj}}} d(k) \ , 
\sum_{k \in \mbox{{\tiny Min}}} d(k)
\right)_{i=1,\ldots,33} \ ,
\end{equation}
where $T_i = \left|k \right|$ is the total number of compositions for composer $i$.
Immediately apparent is Paganini being an outlier to the right, which is due to his almost-exclusive compositions for the violin and guitar, for both of which E/e (degree $\pm4$) is a natural key. 
Likewise, an outlier to the left is Faur\'e, known for his many voice compositions wherein flat keys are natural for singers. 

%%%%%%%%%%%%%%%%%%%%
%%%%%%%%%%%%%%

For reference, the same scatter plot, but with averages weighted by the aforementioned distribution for the combined major/minor compositions over the entire data-set as a normalization, is presented  in Figure \ref{f:avg}(b). 
That is, let the aggregate distribution (both major and minor) at the right of Figure \ref{f:hist}(a), normalized to be a percentage, be $P(k)$, then the 33 points are $(x_i, y_i) = T_i^{-1}
\left(\sum_{k \in \mbox{{\tiny Maj}}} P(k) d(k) \ , 
\sum_{k \in \mbox{{\tiny Min}}} P(k) d(k)\right)$.
This weighting pulls in the extremal points, since higher degrees are weighted less by the distributions.

Simple cluster analysis further reveals interesting groupings. The very bottom triplet of $(7,24,29)$ of Bruckner-Rachmaninoff-Shostakovich explores deep into the minor key. Composers with a large opus for the piano, such as $(7,9,28,26)$, also tend to occupy extremal positions as the instrument allows for easy tonal exploration. Number 11 (Donizetti) is a case where the late Classical composer had a liking for D and A. 

The centroid (by Euclidean distance) is at around $(0.22, -0.31)$, close to Liszt and Mendelssohn who, especially the former, had many transpositions in various keys of earlier composers. The $(4,5,10,17,20,23,25,27)$ group is a solidly Classical/early Romantic set consisting of Mozart, Beethoven and Schubert; the only odd one out temporally is Saint-Sa\"ens, who, though a proponent of his contemporary late Romantics, is indeed known to be compositionally traditional. 
The cluster to the right is interesting, consisting of a mixture of Baroque and (mid-late) Romantic composers. The extremal $(11,26,32,28)$, as mentioned, have an inclination for the keyboard, even more so than $(2,1,30)$, which includes J.~S.~Bach, who, despite the richness of his harmonies, has more grounded key signatures.

%%%%%%%%%%%%%%%%%%%%
\begin{figure}
$
\begin{array}{cc}
(a)
\includegraphics[trim=0mm 0mm 0mm 0mm, clip, width=3in]{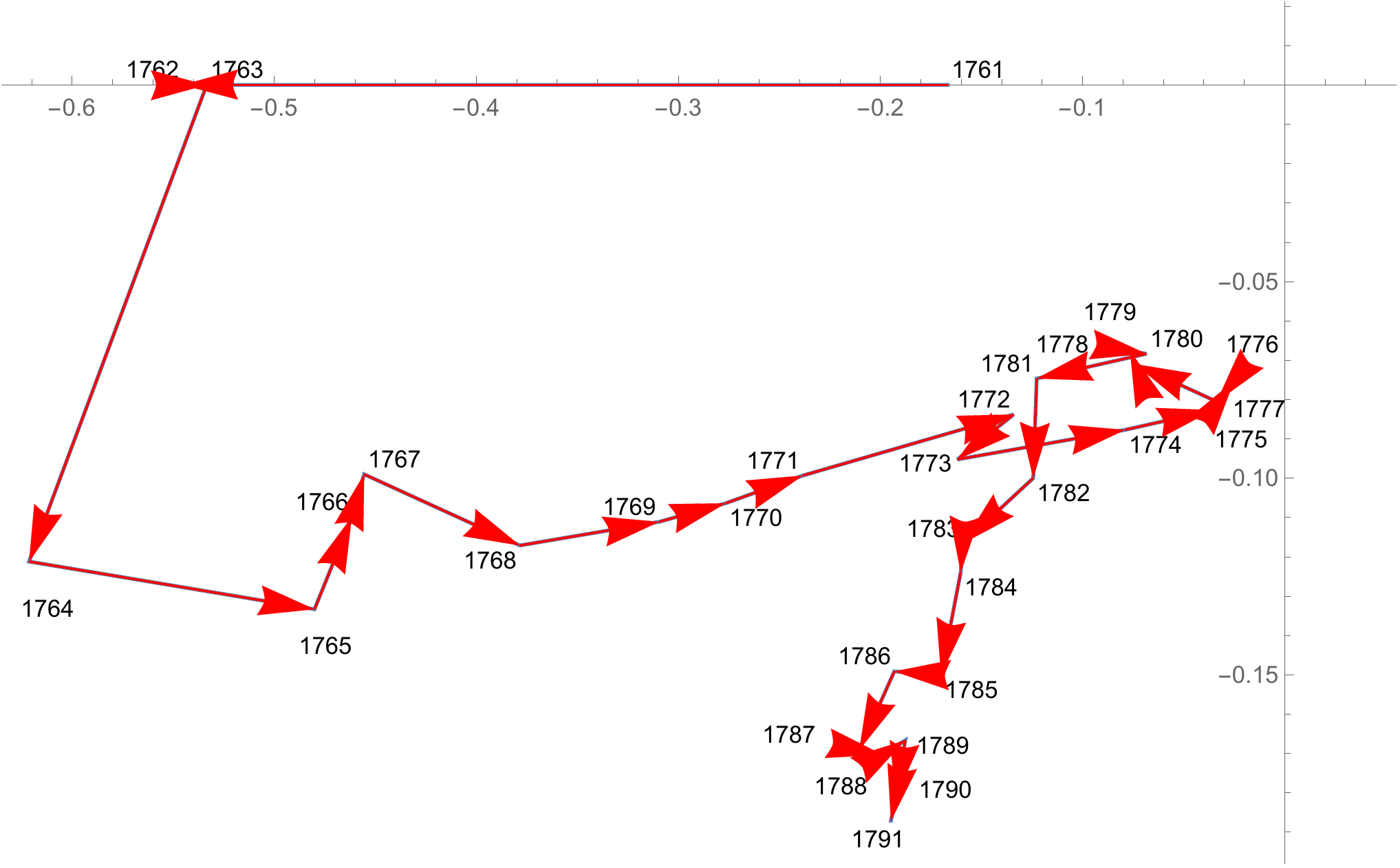}
(b)
\includegraphics[trim=0mm 0mm 0mm 0mm, clip, width=3in]{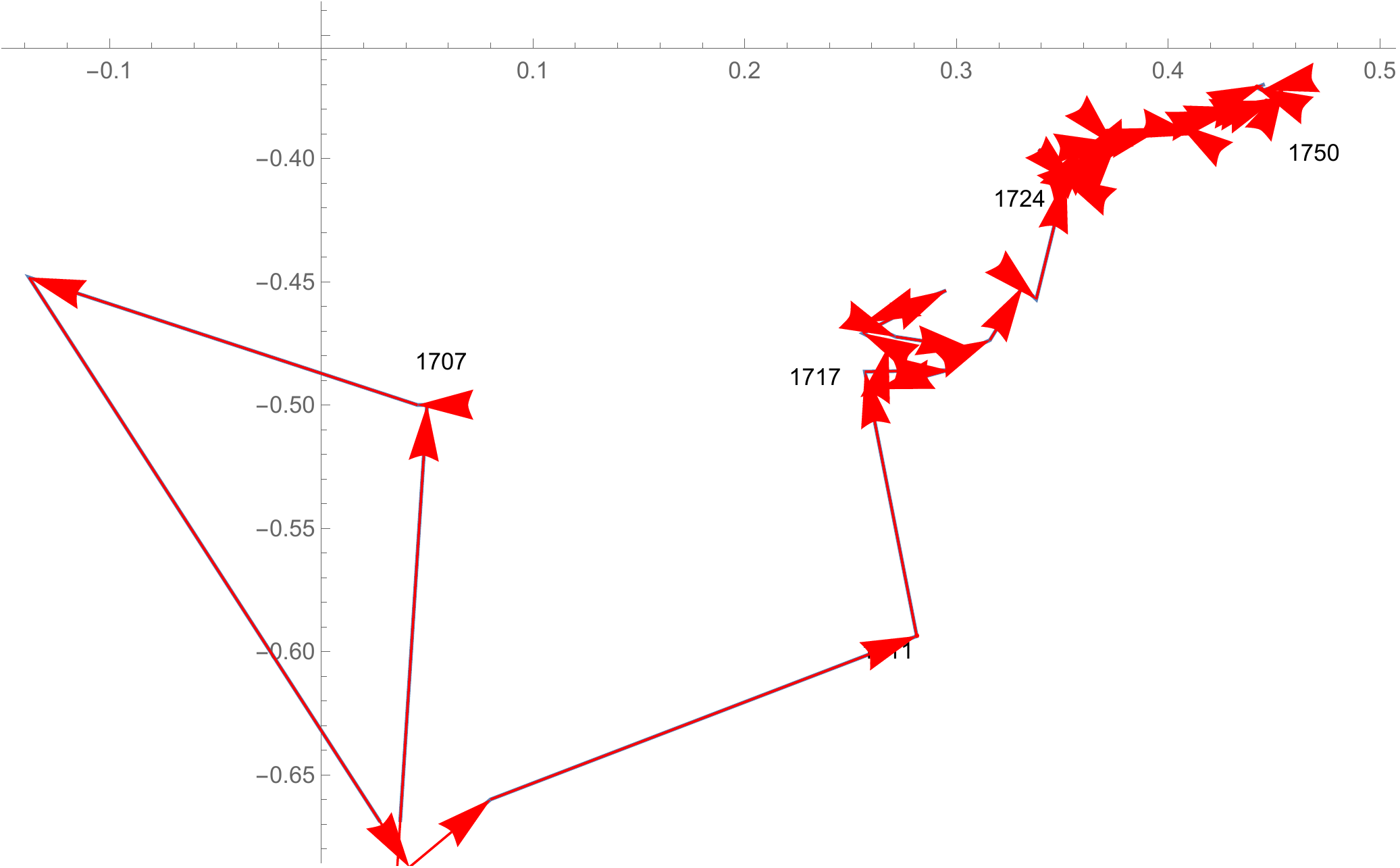}
\end{array}
$
\caption{{\sf
Cumulative average over active years for (a)  W.~A.~Mozart,
(b) J.~S.~Bach on the chromatic diagram.
}
\label{f:life}
}
\end{figure}
%%%%%%%%%%%%%%%%%%%%%%%

%%%%%%%%%%%%%%%%=================
\subsection{Tracking a Particular Career}
The diagram $\cD$ is also useful for tracking a composer's life.
Take Mozart for instance, whose opus is well-catalogued as KV numbers by year \cite{deu,kochel}, a plot of the cumulative arithmetic mean is shown in Figure \ref{f:life}(a). 
The cumulative average is taken because averaging individual years has too many fluctuations. An illustrative trend is seen. In 1761, aged 5, Mozart composed 6 pieces, 3 in C, 2 in 1 in G and 2 in F, giving $(-1/6,0)$. This meandered throughout his career (q.v. \cite{deu}), increasing in compositions in the minor keys, until his Requiem in d (KV626) in 1791. A downward turn towards minors after an upward trend occurred in 1781, when he left Vienna after quarrelling with the archbishop. Similarly, the left turn towards more flat-major keys in 1777 coincided with his departure from Salzburg to Paris, where he encountered new ideas and motifs.

As another example, J.~S.~Bach is shown in Figure \ref{f:life}(b). 
Note that the Bach-Werke-Verzeichnis (BWV) number is not chronological \cite{boyd,sch} thus explicit dates of each composition is obtained in compiling the data and in the analyses.
Again, the biggest fluctuations are in the early years, as the composer was exploring tonally. The trend here, unlike Mozart, is a meandering upward toward Bach's death in 1750, surely due to the fact that he had much less tragic a fate. The step around 1717 approximately corresponds to Bach's leaving Weimar, and the final phase starting with the step around 1723/4 was when he settled down at Leipzig at Thomaskirche till his death in 1750.

\section{Concluding Remarks}
From the various studies, both in the context of an aggregate plot over different composers, as well as in tracking the career progression of individuals, the chromatic diagram has been shown to be a useful indicator. This is rather unexpected since choice of key is a very preliminary factor in a composition; its utility clearly shows how grounded tonality and modality are in the classical tradition. 

This work constitutes only a beginning of much more stylo-statistical studies of musical composition and history; immediate projects which come to mind include plotting the career trends of many more composers, or finding the right degree in going beyond 12-tone compositions. The visual nature of the diagram renders it a particularly powerful tool, whereby making a conducive interplay between music, history and geometry.

%%%=============================================================
\section*{Acknowledgments}
%{\it Catharinae Sanctae Alexandriae adque Majorem Dei Gloriam.}\\
I would like to thank the Science and Technology Facilities Council, UK, for grant ST/J00037X/1, the Chinese Ministry of Education, for a Chang-Jiang Chair Professorship at NanKai University as well as the City of Tian-Jin for a Qian-Ren Award.

{\small
%%
%%%%%%%%%%%%%%%%%%%%%%%%%%%%=======================================

}

\end{document}